\documentclass[preprint,pra,aps,showpacs,preprintnumbers,amsmath,amssymb]{revtex4}

\usepackage{graphicx}
\usepackage{dcolumn}
\usepackage{bm}
\usepackage{subfigure} 
\usepackage{array}
\usepackage{slashbox}
\usepackage{psfrag}
\usepackage{epstopdf}

\begin{document}

\title{Vortex lattices of bosons in deep rotating optical lattices}

\author{Daniel S. Goldbaum}
\email{dsg28@cornell.edu}
\author{Erich J. Mueller}\affiliation{
Laboratory of Atomic and Solid State Physics, Cornell University\\
Ithaca, NY 14853
}

\date{\today}

\begin{abstract}
We study vortex-lattice phases for a Bose gas trapped in a rotating optical-lattice near the superfluid--Mott-insulator transition. We find a series of abrupt structural phase transitions where vortices are pinned with their cores only on plaquettes or only on sites. We discuss connections between these vortex structures and the Hofstadter-butterfly spectrum of free particles on a rotating lattice. 

\end{abstract}

\maketitle
\newcommand{\adi}{\hat{a}_{i}^{\dagger} }
\newcommand{\ai}{\hat{a}_{i} }
\newcommand{\adj}{\hat{a}_{j}^{\dagger} }
\newcommand{\aj}{\hat{a}_{j} }
\newcommand{\numi}{\hat{n}_{i} }

\section{Introduction}
Two of the most exciting directions in cold-atom research involve studying lattice systems and rotating systems \cite{MBPwUCGs}. By increasing the importance of interactions compared to kinetic energy, lattices allow one to study strongly correlated phenomena such as the boson superfluid--Mott-insulator transition \cite{Greiner:2002lr}. These lattice systems are ideal for studying model many-body systems and protocols for quantum information processing \cite{Jaksch:2005lr}. Rotating gases lead to interesting vortex physics \cite{PhysRevLett.84.806, Abo-Shaeer2001, PhysRevLett.89.100403, schweikhard:040404}, and the promise of exotic states such as those which give rise to analogs of fractional quantum-Hall effects \cite{PhysRevLett.84.6, PhysRevLett.87.120405}. Here we study the interplay of lattice physics and rotation physics by calculating the vortex-lattice structures near a Mott transition.

In the absence of an optical lattice a rotating Bose-Einstein Condensate (BEC) develops a triangular lattice of singly quantized vortices \cite{Abo-Shaeer2001, PhysRevLett.89.100403}. This triangular configuration minimizes the logarithmic vortex-vortex interaction. However, as seen in recent experiments far from the Mott regime \cite{tung:240402}, a sufficiently deep optical-lattice potential will pin these vortices at the maxima of that potential \cite{reijnders:060401, pu:190401, reijnders:063607}.

In this paper we show that qualitatively different behavior can be seen in the superfluid state near the Mott-insulator phase. We find that due to changes in the structure of the vortex cores the vortices can actually be pinned at the minima of the potential. In Sec. II we perform numerical mean-field calculations, and find a sequence of first-order transitions between site-centered and plaquette-centered vortex lattices. In Sec. III we use a reduced basis \emph{ansatz} to perform analytic calculations near the Mott boundary, and as a result show how the theory at the Mott boundary is related to the Hofstadter butterfly spectrum. In Sec. IV we summarize our results.  

\section{Numerical calculation of vortex-lattice states}
\subsection{Mean-field theory of the rotating Bose-Hubbard model}
We consider a deep lattice where we can make a tight-binding approximation \cite{PhysRevLett.81.3108}, and the system is described in the rotating frame by a Bose-Hubbard Hamiltonian \cite{wu:043609},
\begin{equation}
\hat{H}_{RBH} = -t \sum_{\langle i,j \rangle} \left( \adi  \aj  \exp{ \left[ i \int_{\bold{r}_j}^{\bold{r}_{i}} d\bold{r} \cdot \bold{A}(\bold{r}) \right]} + H.c. \right) 
+ \frac{1}{2} U \sum_i \numi \left( \numi - 1 \right) - \mu \sum_i \numi \, .
\label{A}
\end{equation}
\newcommand{\tlt}{\tilde{t}}
\newcommand{\tlmu}{\tilde{\mu}}
Above, the operator $\adi $ $\left( \ai \right)$ creates (destroys) a boson and $\numi$ is the number operator at optical-lattice site $i$. The subscript $\langle i,j \rangle $ denotes a nearest-neighbor sum. The parameters $t$, $U$, and $\mu$ are the hopping matrix element, the on-site repulsion strength, and the chemical potential, respectively. Rotation produces the vector potential $\bold{A}(\bold{r}) = \left( m/\hbar\right) \left(\bold{\Omega} \times \bold{r} \right)= \pi \nu \left( x \hat{y} -y \hat{x} \right)$, where $\nu$ is the number of circulation quanta ($h/m$, where m is the atomic mass, and $h$ is $2 \pi$ times Planck's constant $\hbar$) per optical-lattice site. Rotation also produces a harmonic centrifugal-potential which we have assumed is cancelled by a harmonic trap. Although we choose to work in the symmetric gauge our results are not gauge dependent. Scaling energies by U and distances by the lattice constant, the system is characterized by the unitless parameters $\tlt \left( =t/U \right)$, $\tlmu \left(=  \mu/U \right) $ and $\nu$. 

We choose to model a uniform system, rather than explicitly considering a harmonic trap, because we feel that this approach gives more understanding of the phenomena. In addition, we also restrict ourselves to two dimensions, where the physics we are investigating is particularly clear. This geometry can be engineered by applying a sufficiently strong optical lattice in the z-direction which prevents hopping in that direction \cite{DalibardBKT2006}. Also, a rapidly-rotating BEC can assume a similar geometry through centrifugal distortion of its density profile \cite{schweikhard:040404}. Furthermore, we restrict ourselves to the case where the rotation speed is tuned so that $\nu$ is a rational fraction, thus avoiding the commensurability issues which generically occur \cite{tung:240402}. In the strong optical-lattice limit, the vortex lattice will share the geometry of the optical lattice \cite{reijnders:060401, pu:190401, reijnders:063607}.

As one approaches the superfluid-Mott boundary from weak coupling, the vortex cores evolve from empty to containing the Mott phase \cite{wu:043609}. This happens because when the superfluid order is suppressed in the vortex core, the Mott phase is energetically favorable compared to the vacuum. This raises the possibility that the energy of the vortex lattice will be reduced if the cores are centered on optical-lattice minima, ``sites'', rather than at the potential maxima, ``plaquettes''. A competing effect is that if the vortices are site-centered then the overlap of atomic wavepackets centered at neighboring sites will be reduced, raising the kinetic energy. We find that the interplay between these effects leads to a rich structure.

To model an infinite vortex-lattice we perform self-consistent Gutzwiller mean-field calculations on a two-dimensional square-lattice supercell made up of L sites per side, where each site is an optical-lattice potential minimum. We focus on the simplest case where each supercell contains one quantum of circulation, which produces a ground-state solution containing one singly-quantized vortex per supercell, and $\nu = \left( 1/L^2 \right)$. The Gutzwiller mean-field theory can be viewed as a variational calculation where one minimizes $\langle \hat{H}_{RBH} \rangle$ over the Gutzwiller product-states \cite{PhysRevLett.81.3108}, $| \Psi \rangle = \prod_{i} \left( \sum_{n} f_n^i | n \rangle_i \right)$, where $i$ is the site index, $n$ is the particle number, and $| n \rangle_i$ is the $n$-particle occupation-number state at site $i$. Minimizing $\langle \hat{H}_{RBH} \rangle$ with respect to $f_n^{i*}$ with the constraint $\sum_n |f_n^i|^2-1=0$ gives $L^2$ nonlinear eigenvalue equations, one for each site,
\begin{equation}
 -t \sum_{k \text{, nn of } j}  \left( \langle \hat{a}_k \rangle  \sqrt{m}  f_{m-1}^j R_{jk} + \langle \hat{a}_k^{\dagger} \rangle \sqrt{m+1} f_{m+1}^j R_{kj} \right) 
+ \left( \frac{U}{2} m^2- \left( \mu+\frac{U}{2} \right) m + \lambda_j \right) f_m^j = 0 \, ,
\label{M}   
\end{equation}
where the sum is over all nearest neighbors of site $j$, $m$ is the particle-number index, $\lambda_j$ is a Lagrange multiplier, and $R_{jk} = \exp{ \left[ i \int_{\bold{r}_k}^{\bold{r}_{j}} d\bold{r} \cdot \bold{A}(\bold{r}) \right]}$, where $i=\sqrt{-1}$. We iteratively solve these equations: first choosing a trial order-parameter field $\left\{ \alpha_j^{\left( 0 \right)} \right\}$, where $ \alpha_j = \langle \hat{a}_j \rangle $; then updating it by $\alpha_j^{\left( p \right)} = \sum_n \sqrt{n} f_{n-1}^{j*} \left( \left\{ \alpha_j^{\left( p-1 \right)} \right\} \right) f_n^j \left( \left\{ \alpha_j^{\left( p-1 \right)} \right\} \right)$, where $p$ is the iteration index. Similar calculations were performed by Oktel et. al. \cite{oktel:045133} (in a strip geometry) and Wu et. al. \cite{wu:043609} (in a square geometry) to produce vortex lattices.  

We perform calculations in the neighborhood of the $n=1$ Mott phase, so the occupation-number distribution of each site will be peaked about $1$, with small variance. Hence we only need to allow $f_0^j$, $f_1^j$ and $f_2^j$ to be nonzero. In most cases $f_2^j$ and $f_0^j$ will be much smaller than $f_1^j$. We find that using a larger occupation-number basis causes slight shifts of the boundary curves and the energy differences between plaquette- and site-centered vortex-lattice states, but the position of the Mott-lobe is unchanged. To model the infinite vortex lattice with our $\left( L \text{x} L \right)$--supercell we use magnetic boundary conditions \cite{PhysRev.134.A1602, PhysRev.134.A1607}
\begin{eqnarray}
\alpha \left( x+L,y \right) &=& \alpha \left( x, y \right) \exp\left[ -i \frac{\pi}{L} \left( 2 y_0-y \right) \right] \, , \\
\alpha \left( x,y+L \right) &=& \alpha \left( x, y \right) \exp\left[ +i \frac{\pi}{L} \left( 2 x_0-x \right) \right] \, ,
\label{B}
\end{eqnarray}
where $\alpha\left( x,y \right) = \langle \hat{a}_j \rangle$, and $\left( x,y \right)$ are the Cartesian coordinates of site $j$, and $\left( x_0, y_0 \right)$ are free parameters which correspond to the coordinates of the vortex core in our supercell. 

\subsection{Results and discussion}

\begin{figure}
\includegraphics{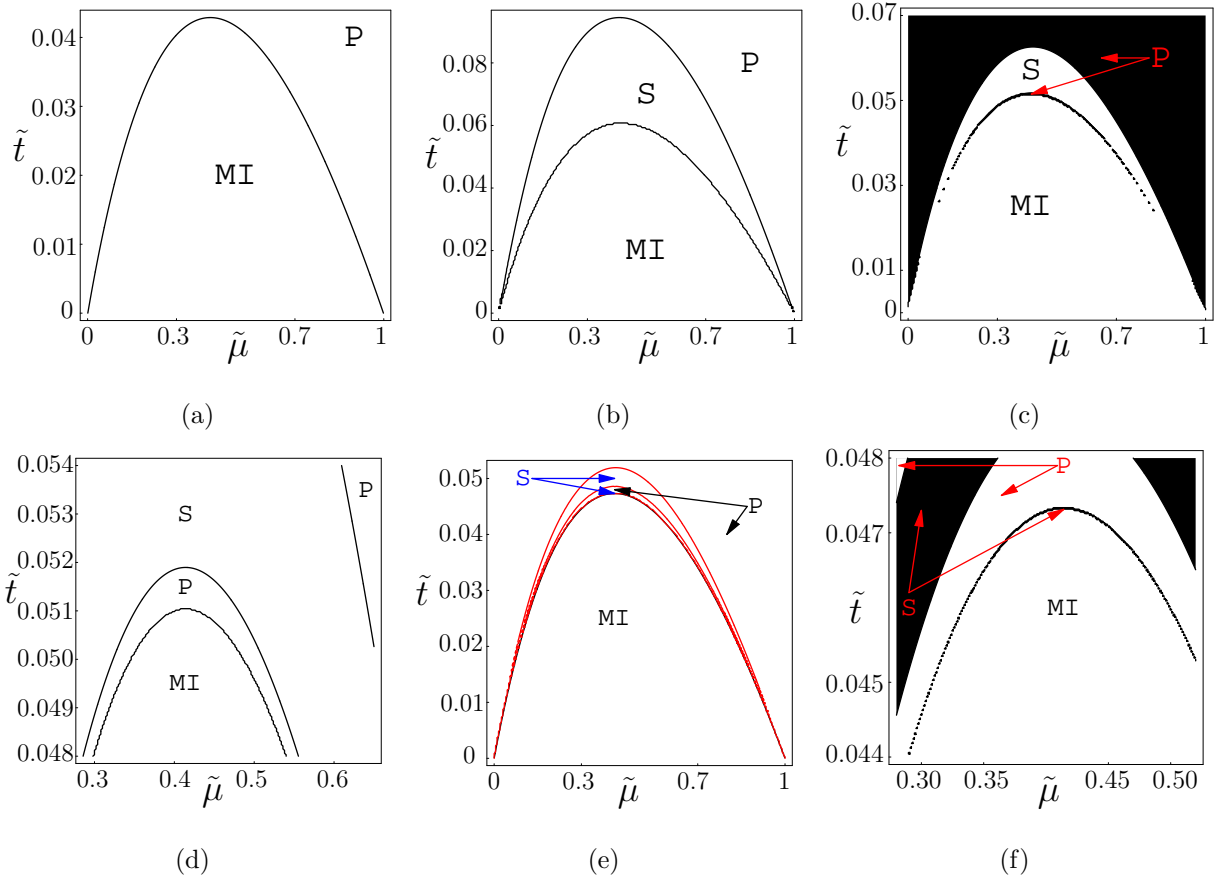}
\caption{(wide, color online) (a)-(b) Structural phase plots for the cases $L=1$ and $L=2$, respectively. Dimensionless parameters $\tilde{t}=t/U$ and $\tilde{\mu}=\mu/U$ represent hopping amplitude and chemical potential, respectively, where each quantity is normalized by the on-site interaction. The plot labels $P$, $S$ and $MI$ refer to P-centered, S-centered and Mott-insulating phases, respectively. (c) The $L=3$ phase plot, where shading is used to emphasize the thin reentrant P phase. (d) A closeup of the critical region of the Mott lobe in (c); the reentrant phase is more clearly resolved. (e) The $L=4$ phase plot, on this parameter range, the inner structural-boundary curve cannot be discerned from the Mott lobe. (f) A closeup of the critical region of the Mott lobe in (e); shading is used to resolve the second reentrant phase region (S phase).
}
\label{2DPhasePlots}
\end{figure}

The phase diagrams for $L=1-4$ are displayed in Fig.~\ref{2DPhasePlots}. Each phase plot has the familiar lobe-shaped Mott-insulator region in the deep-well limit, whose size varies as one changes $n_v$ \cite{PhysRevB.40.546, oktel:045133}. We refer to the plaquette-centered vortex-lattice phase by the symbol $P$, and the site-centered vortex-lattice phase by the symbol $S$. As shown in these phase diagrams we find alternating bands of $P$ and $S$. Moving from weak (large $\tlt$) to strong coupling (small $\tlt$) we find for $L=1$: $P$; $L=2$: $PS$; $L=3$: $PSP$; $L=4$: $PSPS$; $L=5$ (not pictured): $PSPSP$. The bands get very narrow as one increases $L$ and as one approaches the Mott lobe. Table \ref{Bunch} gives the width of the various phases along the line $\tlmu=\tlmu_c$, where $\tlmu_c \left( = \sqrt{2} - 1 \right)$ is the scaled chemical potential at the tip of the $n=1$ Mott lobe.

\begin{table}
\begin{tabular} { | c || c | c | c | c | }
\hline 
\backslashbox{Curve number}{$L =$} & 2 & 3 & 4  \\
\hline \hline
1 & 0.034 & 0.00087 & 0.000017 \\
\hline
2 & ----- & 0.012 &  0.0013 \\
\hline
3 &----- & ----- & 0.0046 \\
\hline
\end{tabular}
\caption{ \label{Bunch} Separation between each structural boundary curve ($L=1-4$) and its corresponding Mott lobe, quantified by $\Delta \, \tlt$ at $\tlmu=\sqrt{2}-1$ (the Mott-lobe tip, see Fig.~\ref{2DPhasePlots}). Curve number $1$ refers to the curve closest to the Mott lobe, curve number $2$ is the next curve out, etc.
 }
\end{table}

There are several important features of these phase diagrams. First, the outermost vortex-lattice phase is always $P$, since a shallow optical-lattice potential pins vortices to the maxima of the potential. Second, for the values of $L$ we have investigated, the phase diagram of a square vortex-lattice configuration characterized by $n_v=1/L^2$ has $L$ phase boundaries. Third, the innermost vortex-lattice phase alternates between $P$ (odd $L$ values) and $S$ (even $L$ values). And finally, the phase boundaries appear to share a universal hyperbolic shape. Although we have no explanation for the second observation, below we will explain the others. 

\begin{figure}
\includegraphics{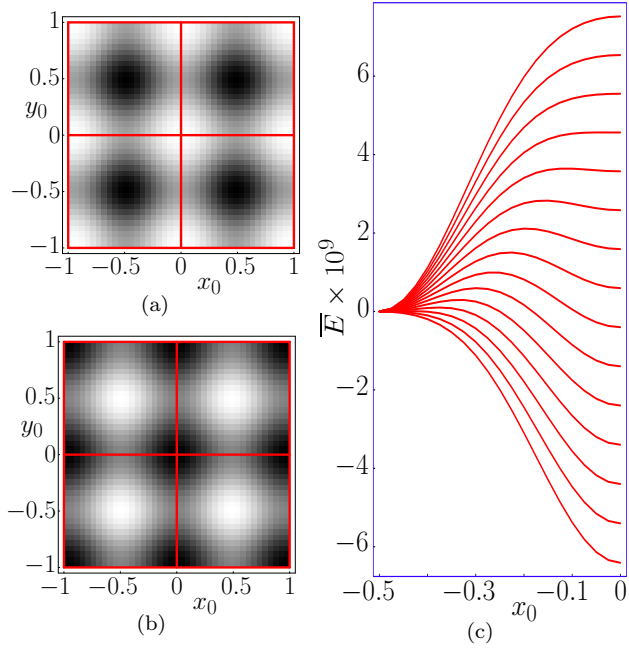}
\caption{(color online)
\emph{Energy} vs \emph{core placement}. Vortex core position $\left( x_0, y_0 \right)$ in units of optical-lattice spacing with $\left( x_0, y_0 \right)=\left( 0,0 \right)$ corresponding to a vortex centered on a site, and $\left( x_0,y_0 \right)=\left( 0.5,0.5 \right)$ corresponding to a vortex centered on a plaquette. These plots correspond to the $L=3$ recurrent phase boundary at $\tlmu=\sqrt{2}-1$, and $ 0.0519 \leq \tlt \leq 0.052$.  In (a) ($\tlt = 0.0519$) and (b) ($\tlt=0.052$) the vertices of the red (gray) lines are sites, and the plots are shaded so that darker (lighter) corresponds to lower (higher) energy. Plot (a) [(b)] corresponds to the P (S) state for $\tlt$ just below (above) the boundary.  (c) A composite of energy vs core-position curves on the diagonal line $y_0=x_0 \in \left( -0.5, 0 \right)$ (from plaquette to site), for $\tlt$ between the spinodal points of the boundary. For each curve $\overline{E}\left( x_0 \right)=\left[ E\left( x_0 \right) - E \left( -0.5 \right)\right]/E_{\textrm{Mott}}$, where $E\left( x_0 \right)= \langle \hat{H}_{RBH} \rangle \left( x_0 \right)$. From top to bottom, this plot has 15 lines corresponding to $\tlt_{\textrm{max}}=0.051902$ and $\tlt_{\textrm{min}}=0.0519015$, with spacing $\Delta \tlt = 7.5 \times 10^{-7}$. } 
\label{EvsCore}
\end{figure}

We analyze the nature of the vortex-configuration phase transition by studying how the energy depends on the location of the vortex core in a single supercell. Figure \ref{EvsCore} illustrates that the transitions are discontinuous. We quantify the abruptness of the phase transition by measuring the width of the coexistence region; that is, we calculate the difference in $\tlt$ (at fixed $\tlmu$) between spinodal points where each of the two energy minima disappear. As shown in Table \ref{Tab1}, the coexistence region becomes thinner as $L$ increases, and as the system moves toward the insulating phase.   

\begin{table}
\begin{tabular} { | c || c | c | c | c | }
\hline 
\backslashbox{Curve number}{$L =$} & 2 & 3 & 4  \\
\hline \hline
1 & 0.014 & $7.5 \times 10^{-6}$ & $ 2.8 \times 10^{-7}$ \\
\hline
2 & ----- & 0.0004 &  $ 1.5 \times 10^{-6}$ \\
\hline
3 &----- & ----- & $2.5 \times 10^{-5}$ \\
\hline
\end{tabular}
\caption{ \label{Tab1} Coexistence region widths, $\Delta \, \tlt$, at $\tlmu=\sqrt{2}-1$ (Mott-lobe tip) for the structural phase boundaries ($L=1-4$). Widths are determined by finding the distance between spinodals. Curve number $1$ refers to the boundary curve closest to the Mott lobe, curve number $2$ is the next curve out, etc. }
\end{table}

\begin{figure}
\includegraphics{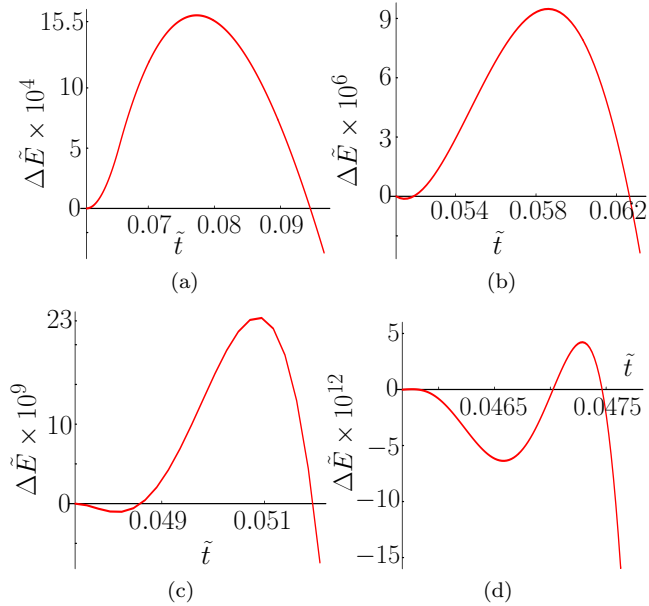}
\caption{(color online)
Energy difference between P and S states with respect to $\tlt$ at fixed values of $L$. (a)-(d) correspond to L=2-5, respectively. The dimensionless energy difference $\Delta \tilde{E} = \left( E_P - E_S \right) / U$, where $E_{P \left( S \right)}= \langle \hat{H}_{RBH} \rangle_{P \left( S \right)}$. The P-centered configuration is always favored in the outermost phase region. The energy differences decrease with decreasing $n_v$ (increasing $L$), and also as the system approaches the insulating region (decreasing $\tlt$). These numbers suggest that, in practice, a homogeneous vortex-lattice configuration is unlikely to be thermally stable in any of the inner phase regions. 
}
\label{EvsT}
\end{figure}

The experimental consequences of our findings depend crucially on the energy difference of the two configurations. For example, the lattice will no longer be pinned if the temperature $T$ exceeds this energy. On the line $\tlmu=\tlmu_c$ we plot these energies in Fig.~\ref{EvsT}.  The pinning energy decreases rapidly as $L$ increases, and as the system approaches the insulating phase in parameter space. We find that the phases inside the outermost $P$ phase all have tiny energy differences. To even see the $L=2$ transition one requires a temperature below $0.15$ nK. Hence our findings are mainly of academic interest. If the temperature is large compared with the pinning energy then the vortex configuration will be determined by the competition between vortex-vortex interaction, which favors a triangular vortex-lattice phase, and entropy, which favors a disordered vortex-liquid.
 
An additional concern is that the structures we find might be in part an artifact of the mean-field theory. Even if this is the case, we believe it is valuable to understand the structure of the mean-field theory. Furthermore, in the following sections we will give arguments which suggest that those results are more general. Finally, we note that experiments are currently far from the regime we consider.

\section{Analytic theory near the Mott-boundary} 

Very near the Mott phase we can linearize Eq. \eqref{A} and analytically calculate the state of the system. During preparation of this paper, Umucahlar and Oktel \cite{umucalilar-2007} presented an independent study with substantial overlap of this section.

\subsection{Reduced-basis \emph{ansatz} and Harper's equation}

It is simplest to illustrate this method by starting with the case of a uniform system which is not rotating $\left( \Omega=0 \right)$. The expectation value of this Hamiltonian with respect to our Gutzwiller product state is
\begin{equation}
\langle \hat{H} \rangle /N = -\sigma \tlt \lvert \alpha \rvert^2 +\frac{1}{2} \langle \hat{n}^2 \rangle - \left( \tlmu + \frac{1}{2} \right) \langle \hat{n} \rangle \, ,
\label{C}
\end{equation}
where $\sigma$ is the number of nearest neighbors, and $N$ is the total number of sites. As one approaches the n-particle Mott lobe we can, as before, make the \emph{ansatz} that the single-site wavefunction is of the form $\lvert \psi \rangle = f_{n-1} \lvert n-1 \rangle + f_n \lvert n \rangle + f_{n+1} \lvert n+1 \rangle$, with $\left( f_{n-1}, f_{n}, f_{n+1} \right) = \left( \epsilon_1, \sqrt{1-\epsilon_1^2-\epsilon_2^2}, \epsilon_2 \right)$, where $\epsilon_i$ are small. One can readily verify that the terms neglected in this \emph{ansatz} are of higher order in $\epsilon$. Minimizing $\langle \hat{H} \rangle$ one finds that the chemical potential at which $\epsilon_i$ becomes nonzero is 
\begin{equation}
\tlmu_{\pm} = \left( n-\frac{1}{2} \left( 1+\sigma \tlt \right) \right) \pm \sqrt{ \frac{1}{4} \sigma^2 \tlt^2 - \left( n+ \frac{1}{2} \right) \sigma \tlt +\frac{1}{4}} \, .
\label{D}
\end{equation} 
In particular, the tip of the Mott Lobe is at $\tlmu_c=\sqrt{n \left( n+1 \right)}-1$, $\sigma \tlt_c= \left[ 2 n + 1 + 2 \sqrt{n \left( n+1 \right)} \right]^{-1}$.

Adding rotation, the energy divided by $U$ is 
\begin{equation}
\langle \hat{H}_{RBH} \rangle = - \sum_{\langle i,j \rangle} \left( \tlt_{ij} \alpha_i^* \alpha_j + c.c. \right) + \sum_i \left( \frac{1}{2} \langle \hat{n}_i^2 \rangle -\left( \tlmu + \frac{1}{2} \right) \langle \hat{n}_i\rangle \right) \, ,
\label{E}
\end{equation}
where $\tlt_{ij}=\tlt \exp{\left[ i \pi \nu \int_{\bold{r}_j}^{\bold{r}_{i}} \left( x \hat{y} -y \hat{x} \right) \cdot d \bold{r} \right]}$. Again, near the Mott lobe we write
\begin{equation}
\left( f_{n-1}^i,f_n^i, f_{n+1}^i \right) = \left( \lambda^i \overline{\alpha}_i^*,\sqrt{1-\lvert \overline{\alpha}_i \rvert^2 \left( \lvert \lambda^i \rvert^2+\lvert \lambda_1^i\rvert^2 \right)},\lambda_1^i \overline{\alpha}_i \right)
\label{F}
\end{equation}
where $\alpha=\overline{\alpha} + O\left( \overline{\alpha}^3 \right)$, and $\lambda_1^i= \frac{1}{\sqrt{n+1}} \left( 1- \sqrt{n} \lambda^i \right)$. Note that unlike our previous calculations, we do not need to restrict $n_v=\nu$.

Next we minimize with respect to $\lambda^i$ to find 
\begin{equation}
\langle \hat{H}_{RBH} \rangle = -\sum_{\langle i,j \rangle} \left( \tlt_{ij} \alpha_i^* \alpha_j + c.c. \right)+ \frac{n-\tlmu}{n+1} \left( 1- n \frac{n-\tlmu}{1+\tlmu} \right) \sum_i \lvert \alpha_i \rvert^2 +E_{\textrm{Mott}} \, ,
\label{G}
\end{equation}
where $E_{\textrm{Mott}}$ is the energy-per-site of the n-particle Mott state, and we have neglected terms of order $\alpha^3$. Next we minimize with respect to $\alpha_k^*$. In the case of the 2D square lattice we arrive at a symmetric-gauge Harper's equation \cite{PhysRevB.14.2239},

\begin{eqnarray}
&-&\alpha \left( x+1,y \right) \exp \left[ i \pi \nu y \right] -\alpha \left( x-1,y \right) \exp \left[ - i \pi \nu y \right] - \alpha \left( x,y+1 \right) \exp \left[ -i \pi \nu x \right] \nonumber \\
&-& \alpha \left( x,y-1 \right) \exp \left[ +i \pi \nu x \right]  + \epsilon \alpha \left( x, y \right) = 0 \, ,
\label{N}
\end{eqnarray}
where
\begin{equation}
\epsilon = \frac{1}{\tlt} \frac{n-\tlmu}{n+1} \left( 1-n \frac{n-\tlmu}{1+\tlmu}\right) \, .
\label{I}
\end{equation}
A simple gauge transformation, $\tilde{\alpha}_j =  \alpha_j  \exp{\left[ -i \pi \nu \int_{\bold{r}_j}^{\bold{r}_{k}} \left( x \hat{y} + y \hat{x} \right) \cdot d \bold{r} \right]}$ along with the assumption that $\tilde{\alpha}\left( x, y \right)= \exp{\left( i \gamma \right)} \beta \left( x \right)$, brings Eq. \eqref{N} into the more familiar form
\begin{equation}
\beta \left( x+1 \right) +\beta \left( x-1 \right) + 2 \cos \left( 2 \pi \nu x - \gamma \right) \beta \left( x \right) = \epsilon \beta \left( x \right) \, ,
\label{M999}
\end{equation}
where the circulation density $\nu=p/q$ is a rational fraction, and $\gamma$ is a wavevector set to $\pi / 2q$ in Ref.~\cite{PhysRevB.14.2239}. The eigenvalue spectrum of Eq. \eqref{N} has an intricate fractal-structure known as the Hofstadter butterfly \cite{PhysRevB.14.2239}. 

\begin{figure}
\includegraphics{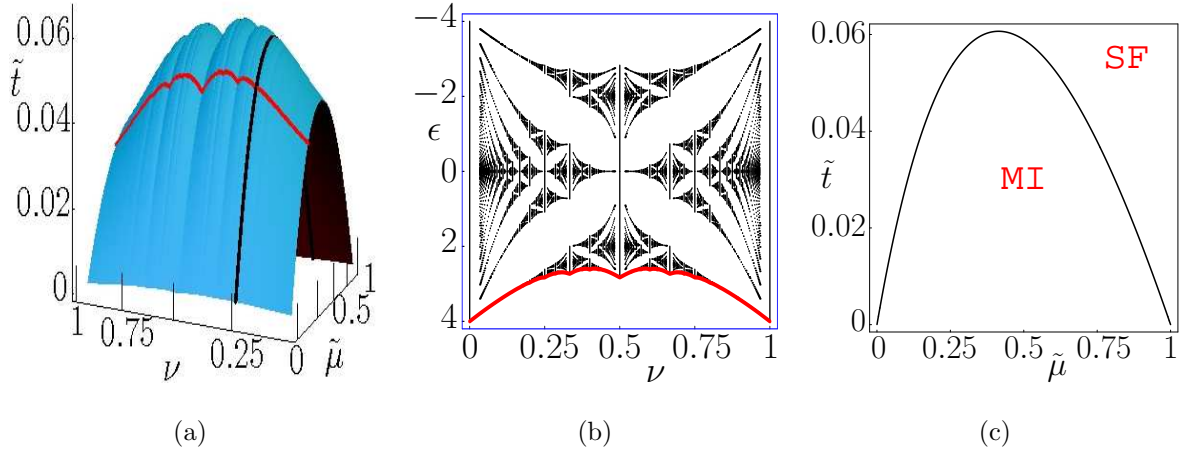}
\caption{(wide, color online)
\label{ButterflyBoundary} 
The blue (light gray) surface in (a) is the mean-field Mott boundary of the Bose-Hubbard model at zero temperature for chemical potential $\tilde{\mu}=\{ 0,1 \}$, and circulation-quanta per optical-lattice site $\nu=\{ 0,1 \}$. The red (dark gray) curve on this surface [and outlining the bottom edge of the spectrum in (b)] demonstrates how, at fixed $\tlmu$ (the value in the figure is $\tlmu=0.2$), the value of $\tlt$ is inversely related to the edge eigenvalues of the Hofstadter butterfly spectrum shown in (b). The black curve on the boundary surface [and in (c)] demonstrates how, at fixed $\nu$ (in this case $\nu=1/4$), the value of $\tlt$ is just a familiar Mott-lobe boundary in the $\left( \tlt ,\tlmu \right)$-plane, as shown in (c). 
}
\end{figure}

Fixing $\nu$ and $\tlmu$, the corresponding point on the Mott lobe is the smallest $\tlt$ for which Eq.~\eqref{I} is an eigenvalue of Eq.~\eqref{N}. This condition is satisifed by the largest eigenvalue $\epsilon=\epsilon_{\textrm{edge}} \left( \nu \right)$ of Eq.~\eqref{N}. We call this largest eigenvalue the \emph{edge eigenvalue}. The Mott boundary is then given by 
\begin{equation}  
\tlt = \frac{1}{\epsilon_{\textrm{edge}} \negthinspace \left[ \nu \right] }  \frac{n-\tlmu}{n+1} \left( 1-n \frac{n-\tlmu}{1+\tlmu}\right) \, ,
\label{K}
\end{equation}
where $n$ is the integer corresponding to the total-particle density in the Mott lobe. This remarkable relationship is illustrated in Fig.~\ref{ButterflyBoundary}. In the non-rotating case we find $\epsilon_{\textrm{edge}} \negthinspace \left[ \nu =0 \right] = 4$, and Eq. \eqref{K} reduces to Eq. \eqref{D}. 

\begin{figure}
\includegraphics{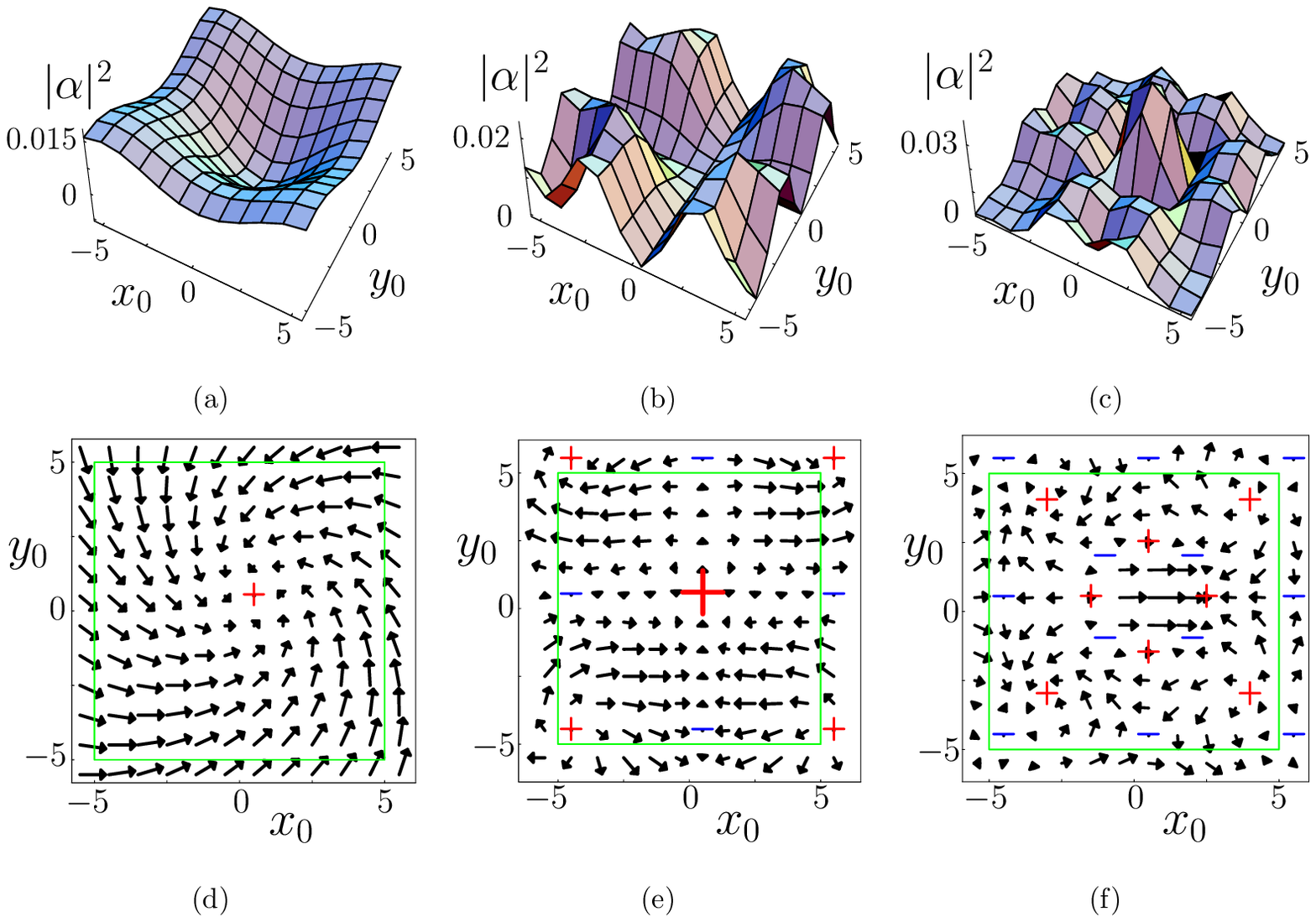}
\caption{(wide, color online)
\label{HBFTextures} \emph{Hofstadter butterfly eigenvectors}, for $\nu=1/100$ in a $10\text{x}10$ supercell. The position coordinates $\left( x_0, y_0 \right)$ are in units of the optical lattice spacing, and the order parameter density $\lvert \alpha \rvert^2$ is normalized so that over a single supercell $\sum_{\left( x_0,y_0 \right)} \lvert \alpha \negmedspace \left( x_0,y_0 \right)\rvert^2=1$. The bands are indexed with $n=1$ for smallest central-eigenvalue, $n=2$ for next smallest, etc. (a)-(c) Plots of order-parameter density $\lvert \alpha \rvert^2$ for bands $n=100$, $n=97$ and $n=91$ respectively.
(d)-(f) The corresponding complex-phase fields. At each site is the base of an arrow pointing in the direction $\left( Re\left[\alpha \right], Im\left[\alpha \right] \right)$, and with length proportional to $\lvert \alpha \rvert$. Positively (negatively) charged vortices are labeled with a red ``$+$'' (blue ``$-$''). The green boundary encloses one unit cell. The size and shape of this boundary are fixed, but varying $\epsilon$ will shift its position. The $n=100$ plot has a single vortex with charge $+1$. The $n=97$ state has a central doubly-quantized vortex of charge $2$, connected by domain walls to vortices of charge $-1$ near the faces of the cell. Vortices of charge $+1$ lie near the corners. The $n=91$ pattern contains 8 ``$+$''--vortices and 7 ``$-$''--vortices in each unit cell.
}
\end{figure}

The eigenvectors of Harper's equation have rich topologies. The highest band (corresponding to the largest $\epsilon$'s) contains states with regular arrays of singly quantized vortices. Changing $\epsilon$ continuously changes the location of the vortices relative to the lattice. The lower bands include states with more complicated structures with multiple vortices of opposite signs. Also, the band structure is symmetric with respect to reflection about $\epsilon=0$. Illustrative structures are shown in Fig.~\ref{HBFTextures}.

The edge state corresponds to an array of singly quantized vortices. For $\nu=1/L^2$ we find that for even or odd $L$ these vortices are site-centered or plaquette-centered, respectively. This explains our previous observation of alternating vortex-lattice phases corresponding to even and odd $L$ values.

\subsection{Discussion}

Why does the Hofstadter butterfly, a pattern associated with noninteracting particles, appear near the Mott lobe, where the interactions are strong? The answer is that near the Mott-lobe boundary most of the atoms are static, with only a dilute gas of mobile particles and/or holes. The diluteness of these excitations leads to single-particle physics.

It should be noted that this explanation does not depend on the approximations of mean-field theory. Even including fluctuations, near the Mott lobe (with the exception of the region immediately about the tip), the system is described by a weakly-interacting gas of excitations \cite{capogrosso-sansone:134302, PhysRevB.40.546}. Thus it is unlikely that the structural transitions we find are an artifact of mean-field theory. Interactions between the excitations can be included in our mean-field theory by including higher-order terms in equation \eqref{G}. If one approximates $\langle \hat{n}_i^2 \rangle = \lvert \alpha_i \rvert^4$, one recovers a nonlinear Schr\"{o}dinger equation  
\begin{equation}
- \sum_{\text{j, nn of k}} \left(   \alpha_j  \exp{\left[ i \pi \nu \int_{\bold{r}_j}^{\bold{r}_{k}} \left( x \hat{y} -y \hat{x} \right) \cdot d \bold{r} \right]} \right) + \lvert \alpha_k \rvert^2 \alpha_k + \frac{\mu}{t} \alpha_k = 0 \, .
\label{P}
\end{equation}

We should mention that one can also study Hofstadter butterfly physics far from the Mott lobe by using a Feshbach resonance \cite{PethickBEC} to tune the interaction of a gas of bosons trapped in a deep, rotating optical lattice. Merely reducing the lattice strength is probably insufficient, as the tight-binding approximation is apt to break down.

\section{Summary}

We have analyzed vortex-lattice phases in a deep optical-lattice potential using the mean-field theory of the rotating Bose-Hubbard Model in a two-dimensional square-lattice at zero temperature. We observed several transitions between site-centered and plaquette-centered vortex states. For the $\left( L \, \text{x} \, L \right)$-supercell calculation (corresponding to $n_v=1/L^2$) there are $L$ boundary curves -- $L-1$ structural curves, and the Mott lobe. We found that the structural transitions are discontinuous, and we quantify trends in the widths of the corresponding coexistence regions as well as trends in the spacing of the structural boundary lines in parameter space. The boundary curves share a universal hyperbolic shape.

We also carried out an analytic study where we determined that the linear eigenvalue equation characterizing the Mott lobe also characterizes the Hofstadter butterfly spectrum. From this we determined an expression for the Mott-lobe boundary. This linearized analysis confirmed the vortex-core placement found in our numerical study.  

\begin{acknowledgments}

The authors would like to acknowledge useful conversations with Kaden R. A. Hazzard. This material is based upon work supported by the National Science Foundation under grant PHY-0456261.

\end{acknowledgments}


\begin{thebibliography}{24}
\expandafter\ifx\csname natexlab\endcsname\relax\def\natexlab#1{#1}\fi
\expandafter\ifx\csname bibnamefont\endcsname\relax
  \def\bibnamefont#1{#1}\fi
\expandafter\ifx\csname bibfnamefont\endcsname\relax
  \def\bibfnamefont#1{#1}\fi
\expandafter\ifx\csname citenamefont\endcsname\relax
  \def\citenamefont#1{#1}\fi
\expandafter\ifx\csname url\endcsname\relax
  \def\url#1{\texttt{#1}}\fi
\expandafter\ifx\csname urlprefix\endcsname\relax\def\urlprefix{URL }\fi
\providecommand{\bibinfo}[2]{#2}
\providecommand{\eprint}[2][]{\url{#2}}

\bibitem[{\citenamefont{Bloch et~al.}(2007)\citenamefont{Bloch, Dalibard, and
  Zwerger}}]{MBPwUCGs}
\bibinfo{author}{\bibfnamefont{I.}~\bibnamefont{Bloch}}, \bibinfo{author}{\bibfnamefont{J.}~\bibnamefont{Dalibard}}, \bibnamefont{and} \bibinfo{author}{\bibfnamefont{W.}~\bibnamefont{Zwerger}}, \emph{\bibinfo{title}{Many-body physics with ultracold gases}}, arXiv:\eprint{0704.3011} (\bibinfo{year}{2007}).

\bibitem[{\citenamefont{Greiner et~al.}(2002)\citenamefont{Greiner, Mandel,
  Esslinger, Hansch, and Bloch}}]{Greiner:2002lr}
\bibinfo{author}{\bibfnamefont{M.}~\bibnamefont{Greiner}},
  \bibinfo{author}{\bibfnamefont{O.}~\bibnamefont{Mandel}},
  \bibinfo{author}{\bibfnamefont{T.}~\bibnamefont{Esslinger}},
  \bibinfo{author}{\bibfnamefont{T.~W.} \bibnamefont{Hansch}},
  \bibnamefont{and} \bibinfo{author}{\bibfnamefont{I.}~\bibnamefont{Bloch}},
  \bibinfo{journal}{Nature} \textbf{\bibinfo{volume}{415}}, \bibinfo{pages}{39}
  (\bibinfo{year}{2002}).

\bibitem[{\citenamefont{Jaksch and Zoller}(2005)}]{Jaksch:2005lr}
\bibinfo{author}{\bibfnamefont{D.}~\bibnamefont{Jaksch}} \bibnamefont{and}
  \bibinfo{author}{\bibfnamefont{P.}~\bibnamefont{Zoller}},
  \bibinfo{journal}{Annals of Physics} \textbf{\bibinfo{volume}{315}},
  \bibinfo{pages}{52} (\bibinfo{year}{2005}).

\bibitem[{\citenamefont{Madison et~al.}(2000)\citenamefont{Madison, Chevy,
  Wohlleben, and Dalibard}}]{PhysRevLett.84.806}
\bibinfo{author}{\bibfnamefont{K.~W.} \bibnamefont{Madison}},
  \bibinfo{author}{\bibfnamefont{F.}~\bibnamefont{Chevy}},
  \bibinfo{author}{\bibfnamefont{W.}~\bibnamefont{Wohlleben}},
  \bibnamefont{and} \bibinfo{author}{\bibfnamefont{J.}~\bibnamefont{Dalibard}},
  \bibinfo{journal}{Phys. Rev. Lett.} \textbf{\bibinfo{volume}{84}},
  \bibinfo{pages}{806} (\bibinfo{year}{2000}).

\bibitem[{\citenamefont{Abo-Shaeer et~al.}(2001)\citenamefont{Abo-Shaeer,
  Raman, Vogels, and Ketterle}}]{Abo-Shaeer2001}
\bibinfo{author}{\bibfnamefont{J.~R.} \bibnamefont{Abo-Shaeer}},
  \bibinfo{author}{\bibfnamefont{C.}~\bibnamefont{Raman}},
  \bibinfo{author}{\bibfnamefont{J.~M.} \bibnamefont{Vogels}},
  \bibnamefont{and} \bibinfo{author}{\bibfnamefont{W.}~\bibnamefont{Ketterle}},
  \bibinfo{journal}{Science} \textbf{\bibinfo{volume}{292}},
  \bibinfo{pages}{476} (\bibinfo{year}{2001}).

\bibitem[{\citenamefont{Engels et~al.}(2002)\citenamefont{Engels, Coddington,
  Haljan, and Cornell}}]{PhysRevLett.89.100403}
\bibinfo{author}{\bibfnamefont{P.}~\bibnamefont{Engels}},
  \bibinfo{author}{\bibfnamefont{I.}~\bibnamefont{Coddington}},
  \bibinfo{author}{\bibfnamefont{P.~C.} \bibnamefont{Haljan}},
  \bibnamefont{and} \bibinfo{author}{\bibfnamefont{E.~A.}
  \bibnamefont{Cornell}}, \bibinfo{journal}{Phys. Rev. Lett.}
  \textbf{\bibinfo{volume}{89}}, \bibinfo{pages}{100403}
  (\bibinfo{year}{2002}).

\bibitem[{\citenamefont{Schweikhard et~al.}(2004)\citenamefont{Schweikhard,
  Coddington, Engels, Mogendorff, and Cornell}}]{schweikhard:040404}
\bibinfo{author}{\bibfnamefont{V.}~\bibnamefont{Schweikhard}},
  \bibinfo{author}{\bibfnamefont{I.}~\bibnamefont{Coddington}},
  \bibinfo{author}{\bibfnamefont{P.}~\bibnamefont{Engels}},
  \bibinfo{author}{\bibfnamefont{V.~P.} \bibnamefont{Mogendorff}},
  \bibnamefont{and} \bibinfo{author}{\bibfnamefont{E.~A.}
  \bibnamefont{Cornell}}, \bibinfo{journal}{Phys. Rev. Lett.}
  \textbf{\bibinfo{volume}{92}}, \bibinfo{eid}{040404} (\bibinfo{year}{2004}).

\bibitem[{\citenamefont{Wilkin and Gunn}(2000)}]{PhysRevLett.84.6}
\bibinfo{author}{\bibfnamefont{N.~K.} \bibnamefont{Wilkin}} \bibnamefont{and}
  \bibinfo{author}{\bibfnamefont{J.~M.~F.} \bibnamefont{Gunn}},
  \bibinfo{journal}{Phys. Rev. Lett.} \textbf{\bibinfo{volume}{84}},
  \bibinfo{pages}{6} (\bibinfo{year}{2000}).

\bibitem[{\citenamefont{Cooper et~al.}(2001)\citenamefont{Cooper, Wilkin, and
  Gunn}}]{PhysRevLett.87.120405}
\bibinfo{author}{\bibfnamefont{N.~R.} \bibnamefont{Cooper}},
  \bibinfo{author}{\bibfnamefont{N.~K.} \bibnamefont{Wilkin}},
  \bibnamefont{and} \bibinfo{author}{\bibfnamefont{J.~M.~F.}
  \bibnamefont{Gunn}}, \bibinfo{journal}{Phys. Rev. Lett.}
  \textbf{\bibinfo{volume}{87}}, \bibinfo{pages}{120405}
  (\bibinfo{year}{2001}).

\bibitem[{\citenamefont{Tung et~al.}(2006)\citenamefont{Tung, Schweikhard, and
  Cornell}}]{tung:240402}
\bibinfo{author}{\bibfnamefont{S.}~\bibnamefont{Tung}},
  \bibinfo{author}{\bibfnamefont{V.}~\bibnamefont{Schweikhard}},
  \bibnamefont{and} \bibinfo{author}{\bibfnamefont{E.~A.}
  \bibnamefont{Cornell}}, \bibinfo{journal}{Phys. Rev. Lett.}
  \textbf{\bibinfo{volume}{97}}, \bibinfo{eid}{240402} (\bibinfo{year}{2006}).

\bibitem[{\citenamefont{Reijnders and Duine}(2004)}]{reijnders:060401}
\bibinfo{author}{\bibfnamefont{J.~W.} \bibnamefont{Reijnders}}
  \bibnamefont{and} \bibinfo{author}{\bibfnamefont{R.~A.} \bibnamefont{Duine}},
  \bibinfo{journal}{Phys. Rev. Lett.} \textbf{\bibinfo{volume}{93}},
  \bibinfo{eid}{060401} (\bibinfo{year}{2004}).

\bibitem[{\citenamefont{Pu et~al.}(2005)\citenamefont{Pu, Baksmaty, Yi, and
  Bigelow}}]{pu:190401}
\bibinfo{author}{\bibfnamefont{H.}~\bibnamefont{Pu}},
  \bibinfo{author}{\bibfnamefont{L.~O.} \bibnamefont{Baksmaty}},
  \bibinfo{author}{\bibfnamefont{S.}~\bibnamefont{Yi}}, \bibnamefont{and}
  \bibinfo{author}{\bibfnamefont{N.~P.} \bibnamefont{Bigelow}},
  \bibinfo{journal}{Phys. Rev. Lett.} \textbf{\bibinfo{volume}{94}},
  \bibinfo{eid}{190401} (\bibinfo{year}{2005}).

\bibitem[{\citenamefont{Reijnders and Duine}(2005)}]{reijnders:063607}
\bibinfo{author}{\bibfnamefont{J.~W.} \bibnamefont{Reijnders}}
  \bibnamefont{and} \bibinfo{author}{\bibfnamefont{R.~A.} \bibnamefont{Duine}},
  \bibinfo{journal}{Phys. Rev. A}
  \textbf{\bibinfo{volume}{71}}, \bibinfo{eid}{063607} (\bibinfo{year}{2005}).

\bibitem[{\citenamefont{Jaksch et~al.}(1998)\citenamefont{Jaksch, Bruder,
  Cirac, Gardiner, and Zoller}}]{PhysRevLett.81.3108}
\bibinfo{author}{\bibfnamefont{D.}~\bibnamefont{Jaksch}},
  \bibinfo{author}{\bibfnamefont{C.}~\bibnamefont{Bruder}},
  \bibinfo{author}{\bibfnamefont{J.~I.} \bibnamefont{Cirac}},
  \bibinfo{author}{\bibfnamefont{C.~W.} \bibnamefont{Gardiner}},
  \bibnamefont{and} \bibinfo{author}{\bibfnamefont{P.}~\bibnamefont{Zoller}},
  \bibinfo{journal}{Phys. Rev. Lett.} \textbf{\bibinfo{volume}{81}},
  \bibinfo{pages}{3108} (\bibinfo{year}{1998}).

\bibitem[{\citenamefont{Wu et~al.}(2004)\citenamefont{Wu, dong Chen, piang Hu,
  and Zhang}}]{wu:043609}
\bibinfo{author}{\bibfnamefont{C.}~\bibnamefont{Wu}},
  \bibinfo{author}{\bibfnamefont{H.}~\bibnamefont{Chen}},
  \bibinfo{author}{\bibfnamefont{J.}~\bibnamefont{Hu}}, \bibnamefont{and}
  \bibinfo{author}{\bibfnamefont{S.} \bibnamefont{Zhang}},
  \bibinfo{journal}{Phys. Rev. A}
  \textbf{\bibinfo{volume}{69}}, \bibinfo{eid}{043609} (\bibinfo{year}{2004}).

\bibitem[{\citenamefont{Hadzibabic et~al.}(2006)\citenamefont{Hadzibabic,
  Kruger, Cheneau, Battelier, and Dalibard}}]{DalibardBKT2006}
\bibinfo{author}{\bibfnamefont{Z.}~\bibnamefont{Hadzibabic}},
  \bibinfo{author}{\bibfnamefont{P.}~\bibnamefont{Kruger}},
  \bibinfo{author}{\bibfnamefont{M.}~\bibnamefont{Cheneau}},
  \bibinfo{author}{\bibfnamefont{B.}~\bibnamefont{Battelier}},
  \bibnamefont{and} \bibinfo{author}{\bibfnamefont{J.}~\bibnamefont{Dalibard}},
  \bibinfo{journal}{Nature} \textbf{\bibinfo{volume}{441}},
  \bibinfo{pages}{1118} (\bibinfo{year}{2006}).

\bibitem[{\citenamefont{Oktel et~al.}(2007)\citenamefont{Oktel, Nita, and
  Tanatar}}]{oktel:045133}
\bibinfo{author}{\bibfnamefont{M.~O.} \bibnamefont{Oktel}},
  \bibinfo{author}{\bibfnamefont{M.}~\bibnamefont{Nita}}, \bibnamefont{and}
  \bibinfo{author}{\bibfnamefont{B.}~\bibnamefont{Tanatar}},
  \bibinfo{journal}{Phys. Rev. B}
  \textbf{\bibinfo{volume}{75}}, \bibinfo{eid}{045133} (\bibinfo{year}{2007}).

\bibitem[{\citenamefont{Zak}(1964{\natexlab{a}})}]{PhysRev.134.A1602}
\bibinfo{author}{\bibfnamefont{J.}~\bibnamefont{Zak}}, \bibinfo{journal}{Phys.
  Rev.} \textbf{\bibinfo{volume}{134}}, \bibinfo{pages}{A1602}
  (\bibinfo{year}{1964}{\natexlab{a}}).

\bibitem[{\citenamefont{Zak}(1964{\natexlab{b}})}]{PhysRev.134.A1607}
\bibinfo{author}{\bibfnamefont{J.}~\bibnamefont{Zak}}, \bibinfo{journal}{Phys.
  Rev.} \textbf{\bibinfo{volume}{134}}, \bibinfo{pages}{A1607}
  (\bibinfo{year}{1964}{\natexlab{b}}).

\bibitem[{\citenamefont{Fisher et~al.}(1989)\citenamefont{Fisher, Weichman,
  Grinstein, and Fisher}}]{PhysRevB.40.546}
\bibinfo{author}{\bibfnamefont{M.~P.~A.} \bibnamefont{Fisher}},
  \bibinfo{author}{\bibfnamefont{P.~B.} \bibnamefont{Weichman}},
  \bibinfo{author}{\bibfnamefont{G.}~\bibnamefont{Grinstein}},
  \bibnamefont{and} \bibinfo{author}{\bibfnamefont{D.~S.}
  \bibnamefont{Fisher}}, \bibinfo{journal}{Phys. Rev. B}
  \textbf{\bibinfo{volume}{40}}, \bibinfo{pages}{546} (\bibinfo{year}{1989}).

\bibitem[{\citenamefont{Umucalilar and Oktel}(2007)}]{umucalilar-2007}
\bibinfo{author}{\bibfnamefont{R.~O.} \bibnamefont{Umucalilar}}
  \bibnamefont{and} \bibinfo{author}{\bibfnamefont{M.~O.} \bibnamefont{Oktel}},
   \bibinfo{journal}{Phys. Rev. A}
  \textbf{\bibinfo{volume}{76}}, \bibinfo{eid}{055601} (\bibinfo{year}{2007}).
  
\bibitem[{\citenamefont{Hofstadter}(1976)}]{PhysRevB.14.2239}
\bibinfo{author}{\bibfnamefont{D.~R.} \bibnamefont{Hofstadter}},
  \bibinfo{journal}{Phys. Rev. B} \textbf{\bibinfo{volume}{14}},
  \bibinfo{pages}{2239} (\bibinfo{year}{1976}).

\bibitem[{\citenamefont{Capogrosso-Sansone
  et~al.}(2007)\citenamefont{Capogrosso-Sansone, Prokof'ev, and
  Svistunov}}]{capogrosso-sansone:134302}
\bibinfo{author}{\bibfnamefont{B.}~\bibnamefont{Capogrosso-Sansone}},
  \bibinfo{author}{\bibfnamefont{N.~V.} \bibnamefont{Prokof'ev}},
  \bibnamefont{and} \bibinfo{author}{\bibfnamefont{B.~V.}
  \bibnamefont{Svistunov}}, \bibinfo{journal}{Phys. Rev. B} \textbf{\bibinfo{volume}{75}},
  \bibinfo{eid}{134302} (\bibinfo{year}{2007}).


\bibitem[{\citenamefont{Pethick and Smith}(2002)}]{PethickBEC}
\bibinfo{author}{\bibfnamefont{C.~J.} \bibnamefont{Pethick}} \bibnamefont{and}
  \bibinfo{author}{\bibfnamefont{H.}~\bibnamefont{Smith}},
  \emph{\bibinfo{title}{Bose-Einstein Condensation in Dilute Gases}}
  (\bibinfo{publisher}{Cambridge University Press}, \bibinfo{year}{2002}).

\end{thebibliography}
\end{document}